\documentclass[12pt,preprint]{aastex}
\usepackage{natbib}

\input epsf

\def\s2n{S^{\prime}/N}

\def\jco{{J=1-0 $^{13}$CO}}
\def\12jco{{J=1-0 $^{12}$CO}}

\begin{document}
\title{The Average Magnetic Field Strength in Molecular Clouds: New Evidence 
of Super--Alfv\'{e}nic Turbulence}

\author{Paolo Padoan\footnote{padoan@jpl.nasa.gov}}
\affil{Department of Physics, University of California, San Diego, La Jolla, CA 92093-0424, USA}
\author{Raul Jimenez}
\affil{Department of Physics and Astronomy, University of Pennsylvania, Philadelphia, PA 19104, USA}
\author{Mika Juvela}
\affil{Helsinki University Observatory, T\"ahtitorninm\"aki, P.O.Box 14, SF-00014 University of Helsinki, Finland}
\author{\AA ke Nordlund}
\affil{Copenhagen Astronomical Observatory and Theoretical Astrophysics Center, DK--2100, Copenhagen, Denmark}

\begin{abstract}

The magnetic field strength in molecular clouds is a fundamental quantity 
for theories of star formation. It is estimated by Zeeman splitting 
measurements in a few dense molecular cores, but its volume--averaged value
within large molecular clouds (over several parsecs) is still uncertain.
In this work we provide a new method to constrain the average magnetic
field strength in molecular clouds.
We compare the power spectrum of gas density of molecular clouds with that 
of two $350^3$ numerical simulations of supersonic MHD turbulence. The 
numerical simulation with approximate equipartition of kinetic and magnetic 
energies (model A) yields the column density power spectrum 
$P(k)\propto k^{-2.25\pm 0.01}$, 
the super--Alfv\'{e}nic simulation (model B) $P(k)\propto k^{-2.71\pm 0.01}$. 
The column density power spectrum of the Perseus, Taurus and Rosetta molecular 
cloud complexes is found to be well approximated by a power law, 
$P_{\rm o}(k)\propto k^{-a}$, with $a=2.74\pm 0.07$, $2.74\pm 0.08$ and 
$2.76\pm 0.08$ respectively. We conclude that the observations are consistent 
with the presence of super--Alfv\'{e}nic turbulence in molecular clouds 
(model B) while model A is inconsistent (more than 99\% confidence) with the observations.

\end{abstract}


\section{Introduction}

The volume--averaged magnetic field strength in molecular clouds has never 
been measured directly. Zeeman splitting has been detected only from a few dense 
molecular cloud cores, where emission lines from molecules such as OH and CN 
are observed \citep{Crutcher99,Bourke+2001}. Dense cores fill only a small 
fraction of the volume of molecular clouds. Therefore, the magnetic
field strength averaged over the molecular cloud volume cannot be directly inferred 
from its value in the cores. This is true especially if the magnetic field strength
has a very intermittent distribution and is correlated with the gas density, as 
suggested by \cite{Padoan+Nordlund99MHD}. 

Estimates of magnetic field strength in molecular clouds have been inferred  
from the dispersion in the polarization angle (e.g. Myers \& Goodman, 1991;
Chrysostomou et al., 1994; Lai et al., 2001; Matthews \& Wilson, 2002; Lai et
al., 2002)
\nocite{Myers+Goodman91,Chrysostomou+94,Lai+2001,Matthews+Wilson2002,Lai+2002} 
as originally suggested by \cite{Davis51} and \cite{Chandrasekhar+Fermi53}.
This method was tested in numerical simulations of MHD turbulence by
 \cite{Ostriker+2001}, \cite{Padoan+2001pol} and \cite{Heitsch+2001}.
The relative motion of ions and neutral molecules, as manifested by a comparison
of their spectral lines, has also been used to estimate the magnetic field strength
in molecular clouds \citep{Houde+2000,Houde+2002}.

In this work we present a new way to constrain the average magnetic field strength
in molecular clouds, based on the density power spectrum. In \S~2 we show  
numerical simulations of supersonic MHD turbulence with different magnetic field 
strength yield different power spectra of gas density. The power spectrum is then
computed from maps of molecular clouds in \S~3. We find that only a rather
low value for the average magnetic field strength, leading to super--Alfv\'{e}nic 
turbulence, is consistent with the observations. Conclusions are summarized in \S~4.

The density power spectrum depends on the rms sonic Mach number of 
the turbulence, $M_{\rm S}$ (the ratio of the rms flow velocity and the sound speed). 
In this work we use only simulations with 
$M_{\rm S}\approx 10$ because that is the approximate value in the molecular
cloud complexes we have studied. We do not study the dependence of the power 
spectrum on the value of $M_{\rm S}$. In general, smaller values of $M_{\rm S}$ yield 
steeper density power spectra than the $M_{\rm S}\approx 10$ models (this was verified
with a set of simulations that will be presented elsewhere). Turbulent flows
with $M_{\rm S} \ll 1$, for example, are expected to generate a Kolmogorov 
density power spectrum, proportional to $k^{-11/3}$. This slope is comparable 
with the electron density power spectrum on very small scale estimated from 
scintillation studies \citep{Armstrong+95}. HI surveys of our galaxy and the 
Magellanic Clouds 
\citep{Crovisier+Dickey83,Elmegreen+2001LMC,Stanimirovic+Lazarian2001,Dickey+2001}
produced power spectra with slope intermediate between the present results 
in molecular clouds and the scintillation studies.

\section{Power Spectrum of Gas Density in Supersonic MHD Turbulence}

In order to study the power spectrum of gas density, we have run two 
simulations of driven supersonic MHD turbulence with rms 
sonic Mach number $M_{\rm S}\approx10$ and isothermal equation of state.
We have solved the three dimensional compressible MHD equations in a 
staggered mesh with 350$^3$ computational cells and periodic boundary 
conditions. The initial magnetic and density fields are uniform. 
The flow is driven by an external large scale random and solenoidal force, 
correlated at the largest scale turn--over time. The time derivative of the 
random force is generated in Fourier space, with power only in the range of 
wavenumbers $1\le k\, L_{\rm mesh}/ 2\,\pi\le 2$ . The initial velocity field 
is proportional to the initial force, with an rms amplitude of approximately
50\% of its relaxed value. Details about the numerical method are given in 
\cite{Padoan+Nordlund99MHD}.  

We have run the simulations for five dynamical times. The dynamical time is 
here defined as $t_{\rm d}=L_{\rm mesh}/(2\, u)$, where 
$u$ is the rms flow velocity. We choose this definition of 
dynamical time because the flow is forced up to wavenumber $k=4\pi/L_{\rm mesh}$ 
and therefore the largest turbulent scale is approximately $L_{\rm mesh}/2$.  

We characterize the simulations based on the relative importance of magnetic
and dynamic (turbulent) pressure. The pressure ratio is defined as
$P_{\rm m}/P_{\rm d}= \langle B^2\rangle / [8\pi\,\langle \rho u^2\rangle]$,
averaging over the computational volume and over the last four dynamical 
times. In the first simulation (model A) the statistically relaxed pressure ratio is 
$P_{\rm m}/P_{\rm d}\approx 0.65\pm0.05$ (approximate equipartition of magnetic 
and kinetic energies). In the second simulation (model B), the ratio is 
$P_{\rm m}/P_{\rm d}\approx 0.09\pm0.01$ (kinetic energy of the turbulence
approximately five times larger than magnetic energy). 
The standard deviation of approximately 10\% in the pressure
ratio of models A and B is due primarily to time fluctuations of the random
force. 

\begin{figure}
\plotone{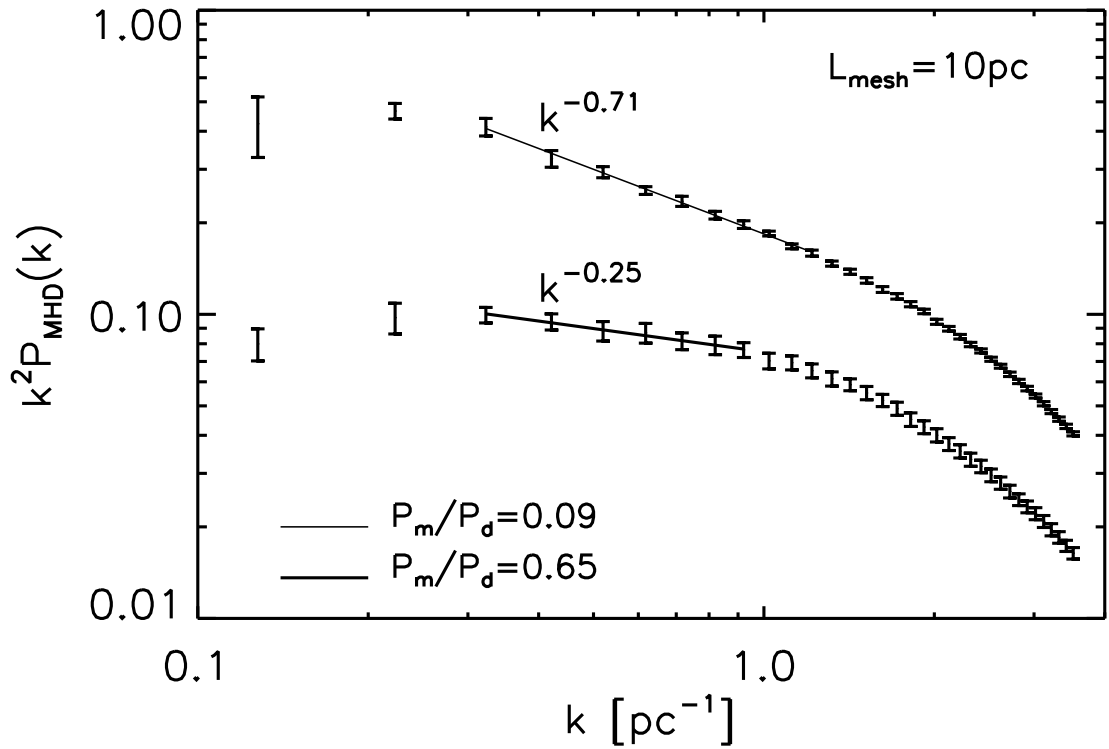}
\figcaption{Time--averaged density power spectra of model B (upper plot),
and model A (lower plot). The solid lines are least 
square fits in the range 0.3--0.9~pc$^{-1}$ for model A and 0.3--1.2~pc$^{-1}$ 
for model B, assuming a mesh size $L_{\rm mesh}=10$~pc.}
\label{fig2}
\end{figure}

Power spectra have been computed for 18 times over the last four dynamical times
(the small scale portions of these power spectra are statistically independent
because the dynamical time decreases with spatial scale).     
The slopes are computed from a least square fit of the time--averaged
power spectra plotted in Figure~\ref{fig2}.
We find that the power spectrum of the density field is sensitive to the pressure 
ratio (or the average magnetic field strength). The power spectrum of the 
three dimensional (3D) density field is $P(k)\propto k^{-2.25\pm 0.01}$ in the 
equipartition run (model A) and $P(k)\propto k^{-2.70\pm 0.01}$ in the super--Alfv\'{e}nic 
case (model B). 

In isotropic turbulence the power spectrum of the projected density is the same 
as the power spectrum of the 3D density field (not necessarily in the presence
of a mean magnetic field or in real molecular clouds). We have verified 
this in our numerical data. We have also verified that the power spectrum of the
projected density does not depend on the direction of projection relative to the 
direction of the mean magnetic field. However, our results are based on the power 
spectra of the 3D density field and not of the projected density, because the 
statistical sample size is reduced by the projection (larger noise in 2D power 
spectra than in 3D ones).

\section{Power Spectrum of Column Density in Molecular Clouds}

The distribution of column density in molecular clouds can be estimated from 
maps of the \jco\ emission line. In this work we use \jco\ maps of the Perseus 
\citep{Padoan+99per}, Taurus \citep{Mizuno+95} and Rosetta \citep{Heyer+2003} 
molecular cloud complexes and find the power spectrum of projected density
in these regions is well approximated by a power law, 
$P_{\rm o}(k)\propto k^{-a_{\rm o}}$. 

The value of the gas column density inferred from the \jco\ maps
depends on the distribution of $^{13}$CO abundance and \jco\ 
excitation temperature, $T_{\rm ex}$. The column density can be 
estimated using the LTE method \citep{Dickman78,Harjunpa+Mattila96,Padoan+98co}.
In the LTE method the value of $T_{\rm ex}$ along each line of sight is
assumed to be constant and is estimated using the observed peak temperature, 
$T_{\rm r}=T_{\rm p}$, of an optically thick line ($\tau \gg 1$) in the equation:
\begin{equation}
T_{\rm r}=[J(T_{\rm ex})-J(T_{\rm bg})](1-e^{-\tau})
\label{for1}
\end{equation}
where $T_{\rm bg}=2.7$~K is the background temperature, and the function 
$J(T)$ is defined as:
\begin{equation}
J(T)=\frac{T_0}{\exp(T_0/T)-1}
\label{for2}
\end{equation}
with $T_0=h\nu_{\rm 10}/k$ and $\nu_{\rm 10}$ is the frequency of the 
\jco\ transition. The \12jco\ transition, when available, is generally 
used as the optically thick line to estimate the value of $T_{\rm ex}$.
For the present analysis we apply the LTE method using only the
\jco\ line because $^{12}$CO maps are not available to us. 
This line is optically thick only in the brightest 
regions of the maps used in the present work. The value of 
$T_{\rm ex}$ can be estimated from the \jco\ line only in those regions.
Therefore, we use the peak temperature of this line over the whole map 
to estimate the value of $T_{\rm ex}$ as outlined above. This single value 
of $T_{\rm ex}$ is used for all map positions (the effect of this assumption
is addressed below with radiative transfer calculations). 
Finally, the $^{13}$CO column density is given by:
\begin{equation}
N_{\rm LTE}=
6.39\times 10^{14}\,Q\,\frac{\Sigma_v{\tau(v)\Delta v}}{1-e^{-T_0/T_{\rm ex}}}
\label{for3}
\end{equation}
where $\tau(v)$ is the optical depth in the velocity channel 
corresponding to the velocity $v$ and is estimated from equation
(\ref{for1}) using the estimated value of $T_{\rm ex}$ and the radiation
temperature $T_{\rm r}(v)$ given by the observed line profile. $\Delta v$
is the width of the velocity channels in the observations, expressed in km/s, 
and the partition function $Q$ is assumed to be a constant over the map so
its value is not required as it does not affect the slope of the density power
spectrum. 

\begin{figure}
\plotone{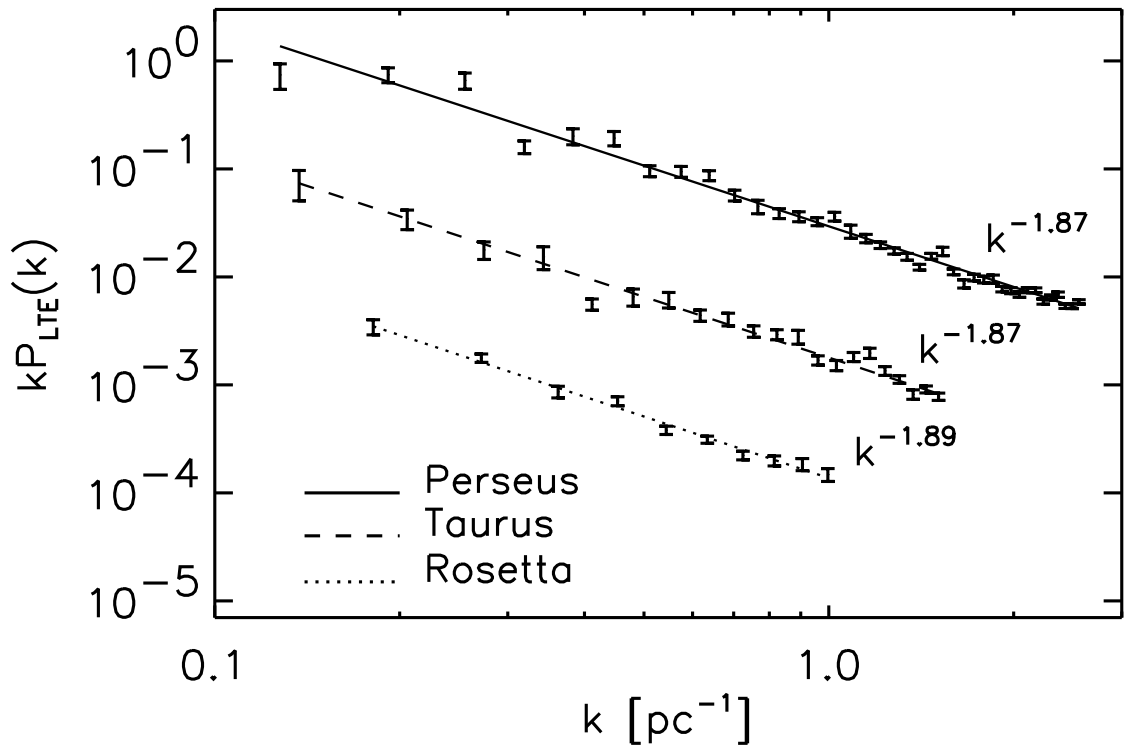}
\figcaption{Power spectra of three molecular cloud complexes.
The power spectra are computed from images of column density obtained with 
the LTE method applied to maps of the \jco\ emission line.}
\label{fig3}
\end{figure}

Column density maps of the Perseus, Taurus and Rosetta regions have been 
computed from the observed \jco\ spectral maps with this simplified LTE method. 
Their power spectra are plotted in Figure~\ref{fig3}. They are shown to be well 
approximated by power laws over one decade or more in wavenumber, 
$P_{\rm o}(k)\propto k^{-a_{\rm o}}$, with $a_{\rm o}\approx 2.87\pm 0.04$ for Perseus, 
$a_{\rm o}\approx 2.87\pm 0.06$ for Taurus and $a_{\rm o}\approx 2.89\pm 0.06$
for Rosetta. 

In order to estimate the effect of assuming constant $Q$ and $T_{\rm ex}$, 
the simplified LTE method for estimating the gas column density has been 
applied to synthetic spectra. Assuming a uniform $^{13}$CO abundance, 
synthetic spectral maps of the \jco\ line have been
computed with a 3D non--LTE Monte Carlo radiative transfer code \citep{Juvela97} 
as described in \cite{Padoan+98cat}. The radiative transfer is computed through the
density and velocity field of two snapshots of model B (at 3 and 4 dynamical times), 
regridded to a resolution
of $175^3$, assuming average density  $\langle n \rangle=500$~cm$^{-3}$ and 
mesh size $L_{\rm mesh}=10$~pc. The 3D distribution of kinetic temperature is 
computed self--consistently as part of the radiative transfer solution, from 
the balance of cosmic ray heating and molecular and atomic cooling. 
We have assumed a cosmic ray hydrogen ionization rate of $2\times 10 ^{-17}$~s$^{-1}$
and heating of 8~eV per ionization.

\begin{figure}
\plotone{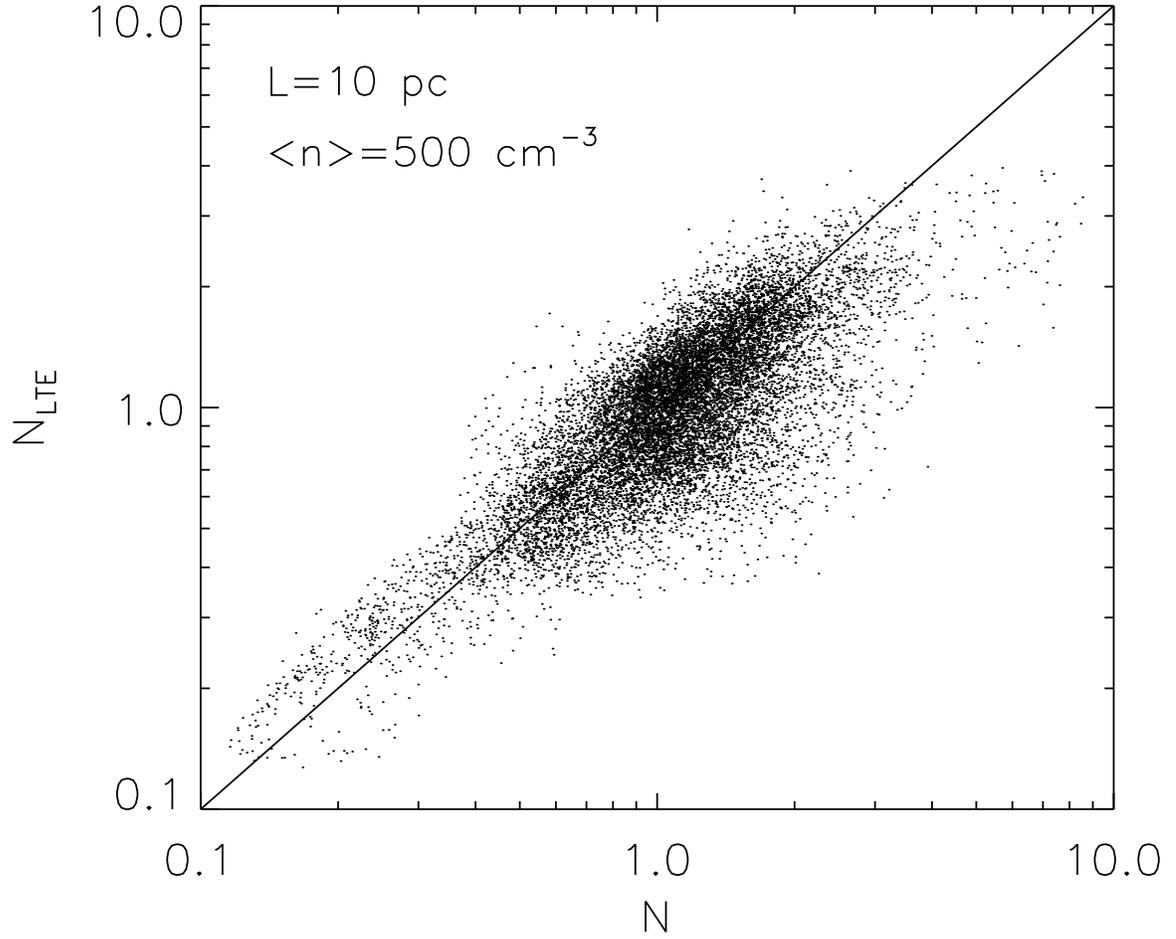}
\figcaption{Scatter plot of column density estimated with the
LTE method applied to a synthetic map of the \jco\ emission line versus
the column density in the MHD data cube used to compute the synthetic spectra.}
\label{fig4}
\end{figure}

Six maps of $175\times175$ synthetic \jco\ lines are obtained in this way,
corresponding to three orthogonal directions of projection of the 
two 3D snapshots. The simplified LTE method is then applied to these six 
synthetic spectral maps and the estimated column density is compared with the 
actual column density in the original snapshots of model B. The result for 
one of the maps is plotted in Figure~\ref{fig4}. Figure~\ref{fig4} shows the 
estimated column density, $N_{\rm LTE}$, is roughly proportional to the true 
column density, but tends to saturate at large column density values. 
This is due both to the low gas temperature and to the \jco\ line saturation 
in the densest regions. The power spectra of the column
density field estimated from the synthetic spectral map, 
$P_{\rm LTE}(k)\propto k^{-a_{\rm LTE}}$, 
are then compared with the power spectra computed directly from the 
projections of the original MHD data cube,
$P_{\rm MHD}(k)\propto k^{-a_{\rm MHD}}$. 

We find that $P_{\rm LTE}(k)$ is generally steeper than $P_{\rm MHD}(k)$.
This is primarily due to the decreased emission from the densest regions
due to their low gas temperature and to the saturation of the \jco\ line. 
The difference is 
$a_{\rm LTE}-a_{\rm MHD}=0.13\pm 0.06$. If this correction is applied to
the observed molecular cloud complexes, the corrected power spectrum 
has a slope $a=2.74\pm 0.07$ for Perseus, $a=2.74\pm 0.08$ for Taurus and
$a=2.76\pm 0.08$ for .
At the 1 $\sigma$ level, model B is consistent with these molecular cloud
complex power spectra, while the power spectrum of model A is 
inconsistent with the observations with virtually 100\% confidence 
(7 $\sigma$).

Variations in $^{13}$CO abundance may affect the estimated column density.
However, significant CO depletion is expected only above 10 magnitudes of visual 
extinction and the CO abundance should drop only below 1-2 magnitudes. Most of
the gas mass in molecular cloud complexes emits at values of visual extinction 
between 1 and 10 mag. Therefore, the effect of gas temperature variations 
estimated above should be more important than the neglected effect of variations 
in $^{13}$CO abundance.

\section{Conclusions}

The average magnetic field strength in molecular clouds cannot be
measured directly. However, it can be inferred from observational
data due to its effect on the gas dynamics. In this work we have 
found the power spectrum of the density field is a sensitive 
diagnostic of the magnetic field strength. Numerical simulations
of supersonic MHD turbulence with rms sonic Mach number 
$M_{\rm S}\approx 10$ develop a power law power spectrum of gas
density, $P_{\rm MHD}(k)\propto k^{-a_{\rm MHD}}$. The value
of the power law exponent is $a_{\rm MHD}=2.25\pm 0.01$ when
the flow rms velocity is comparable to the Alfv\'{e}n velocity
and $a_{\rm MHD}=2.71\pm 0.01$ in the super--Alfv\'{e}nic simulation.

The density power spectrum can be measured also in molecular cloud
complexes, for example using \jco\ maps. However, the column density
(and its power spectrum) estimated using only the \jco\ line and the 
LTE method are biased by a number of uncertainties. The most significant 
uncertainties are the 3D distribution of the gas kinetic temperature in the 
molecular cloud complexes and the saturation of the \jco\ line in very dense 
regions. We have investigated the effect of these uncertainties 
in the density power spectrum using synthetic 
\jco\ spectral maps computed with a non--LTE Monte Carlo radiative transfer
code. The 3D equilibrium temperature distribution is computed self--consistently
as part of the radiative transfer solution by balancing cosmic ray heating
with molecular and atomic cooling. The correct power spectrum slope, 
$a_{\rm MHD}$, is found to be smaller than the slope estimated with the 
LTE method, $a_{\rm LTE}$, with $a_{\rm MHD}=a_{\rm LTE}-0.13\pm 0.06$.

With this correction, the power spectrum slope is $a=2.74\pm 0.07$ for
Perseus,$a=2.74\pm 0.08$  for Taurus and $a=2.76\pm 0.08$ for Rosetta.
The super--Alfv\'{e}nic model is consistent with this result, while
the model with rms flow velocity comparable to the Alfv\'{e}n velocity
is ruled out by the observations. This is yet another indication that 
super--Alfv\'{e}nic turbulence provides a good description 
of molecular cloud dynamics. As first proposed by
\cite{Padoan+Nordlund97MHD,Padoan+Nordlund99MHD}, the average magnetic field
strength in molecular clouds may be much smaller than required to support
them against the gravitational collapse. Evidence of super--Alfv\'{e}nic
turbulence was also found recently by Troland and Heiles (2001) in HI clouds. 

\nocite{Troland+Heiles2001}

\acknowledgements

MJ acknowledges the support from the Academy of Finland Grants no.
174854 and 175068. Computing time was provided by the Danish Center 
for Scientific Computing. The research of RJ is partially supported by NSF grant AST-0206031.



\begin{thebibliography}{30}
\expandafter\ifx\csname natexlab\endcsname\relax\def\natexlab#1{#1}\fi

\bibitem[{{Armstrong} {et~al.}(1995){Armstrong}, {Rickett}, \&
  {Spangler}}]{Armstrong+95}
{Armstrong}, J.~W., {Rickett}, B.~J., \& {Spangler}, S.~R. 1995, ApJ, 443, 209

\bibitem[{Bourke {et~al.}(2001)Bourke, Myers, Robinson, \&
  Hyland}]{Bourke+2001}
Bourke, T., Myers, P., Robinson, G., \& Hyland, H. 2001, ApJ

\bibitem[{{Chandrasekhar} \& {Fermi}(1953)}]{Chandrasekhar+Fermi53}
{Chandrasekhar}, S. \& {Fermi}, E. 1953, ApJ, 118, 113

\bibitem[{{Chrysostomou} {et~al.}(1994){Chrysostomou}, {Hough}, {Burton}, \&
  {Tamura}}]{Chrysostomou+94}
{Chrysostomou}, A., {Hough}, J.~H., {Burton}, M.~G., \& {Tamura}, M. 1994,
  MNRAAS, 268, 325

\bibitem[{{Crovisier} \& {Dickey}(1983)}]{Crovisier+Dickey83}
{Crovisier}, J. \& {Dickey}, J.~M. 1983, \aap, 122, 282

\bibitem[{{Crutcher}(1999)}]{Crutcher99}
{Crutcher}, R.~M. 1999, ApJ, 520, 706

\bibitem[{{Davis}(1951)}]{Davis51}
{Davis}, L. 1951, Physical Review, 81, 890

\bibitem[{{Dickey} {et~al.}(2001){Dickey}, {McClure-Griffiths}, {Stanimirovi{\'
  c}}, {Gaensler}, \& {Green}}]{Dickey+2001}
{Dickey}, J.~M., {McClure-Griffiths}, N.~M., {Stanimirovi{\' c}}, S.,
  {Gaensler}, B.~M., \& {Green}, A.~J. 2001, ApJ, 561, 264

\bibitem[{Dickman(1978)}]{Dickman78}
Dickman, R.~L. 1978, \apjs, 37, 407

\bibitem[{{Elmegreen} {et~al.}(2001){Elmegreen}, {Kim}, \&
  {Staveley-Smith}}]{Elmegreen+2001LMC}
{Elmegreen}, B.~G., {Kim}, S., \& {Staveley-Smith}, L. 2001, ApJ, 548, 749

\bibitem[{Harjunp\"{a}\"{a} \& Mattila(1996)}]{Harjunpa+Mattila96}
Harjunp\"{a}\"{a}, P. \& Mattila, K. 1996, \aap, 305, 920

\bibitem[{{Heitsch} {et~al.}(2001){Heitsch}, {Zweibel}, {Mac Low}, {Li}, \&
  {Norman}}]{Heitsch+2001}
{Heitsch}, F., {Zweibel}, E.~G., {Mac Low}, M., {Li}, P., \& {Norman}, M.~L.
  2001, ApJ, 561, 800

\bibitem[{Heyer \& et~al.(2003)}]{Heyer+2003}
Heyer, M.~H. \& et~al. 2003, in preparation

\bibitem[{{Houde} {et~al.}(2002){Houde}, {Bastien}, {Dotson}, {Dowell},
  {Hildebrand}, {Peng}, {Phillips}, {Vaillancourt}, \& {Yoshida}}]{Houde+2002}
{Houde}, M., {Bastien}, P., {Dotson}, J.~L., {Dowell}, C.~D., {Hildebrand},
  R.~H., {Peng}, R., {Phillips}, T.~G., {Vaillancourt}, J.~E., \& {Yoshida}, H.
  2002, ApJ, 569, 803

\bibitem[{{Houde} {et~al.}(2000){Houde}, {Bastien}, {Peng}, {Phillips}, \&
  {Yoshida}}]{Houde+2000}
{Houde}, M., {Bastien}, P., {Peng}, R., {Phillips}, T.~G., \& {Yoshida}, H.
  2000, \apj, 536, 857

\bibitem[{Juvela(1997)}]{Juvela97}
Juvela, M. 1997, A\& A, 322, 943

\bibitem[{{Lai} {et~al.}(2001){Lai}, {Crutcher}, {Girart}, \& {Rao}}]{Lai+2001}
{Lai}, S., {Crutcher}, R.~M., {Girart}, J.~M., \& {Rao}, R. 2001, ApJ, 561, 864

\bibitem[{{Lai} {et~al.}(2002){Lai}, {Crutcher}, {Girart}, \& {Rao}}]{Lai+2002}
---. 2002, ApJ, 566, 925

\bibitem[{{Matthews} \& {Wilson}(2002)}]{Matthews+Wilson2002}
{Matthews}, B.~C. \& {Wilson}, C.~D. 2002, ApJ, 574, 822

\bibitem[{{Mizuno} {et~al.}(1995){Mizuno}, {Onishi}, {Yonekura}, {Nagahama},
  {Ogawa}, \& {Fukui}}]{Mizuno+95}
{Mizuno}, A., {Onishi}, T., {Yonekura}, Y., {Nagahama}, T., {Ogawa}, H., \&
  {Fukui}, Y. 1995, ApJL, 445, L161

\bibitem[{{Myers} \& {Goodman}(1991)}]{Myers+Goodman91}
{Myers}, P.~C. \& {Goodman}, A.~A. 1991, ApJ, 373, 509

\bibitem[{{Ostriker} {et~al.}(2001){Ostriker}, {Stone}, \&
  {Gammie}}]{Ostriker+2001}
{Ostriker}, E.~C., {Stone}, J.~M., \& {Gammie}, C.~F. 2001, ApJ, 546, 980

\bibitem[{Padoan {et~al.}(1999)Padoan, Bally, Billawala, Juvela, \&
  Nordlund}]{Padoan+99per}
Padoan, P., Bally, J., Billawala, Y., Juvela, M., \& Nordlund, {\AA}. 1999,
  ApJ, 525, 318

\bibitem[{{Padoan} {et~al.}(2001){Padoan}, {Goodman}, {Draine}, {Juvela},
  {Nordlund}, \& {R{\" o}gnvaldsson}}]{Padoan+2001pol}
{Padoan}, P., {Goodman}, A., {Draine}, B.~T., {Juvela}, M., {Nordlund}, {\AA}.,
  \& {R{\" o}gnvaldsson}, {\" O}.~E. 2001, \apj, 559, 1005

\bibitem[{Padoan {et~al.}(1998{\natexlab{a}})Padoan, Juvela, Bally, \&
  Nordlund}]{Padoan+98co}
Padoan, P., Juvela, M., Bally, J., \& Nordlund, {\AA}. 1998{\natexlab{a}}, ApJ,
  in press

\bibitem[{Padoan {et~al.}(1998{\natexlab{b}})Padoan, Juvela, Bally, \&
  Nordlund}]{Padoan+98cat}
---. 1998{\natexlab{b}}, ApJ, 504, 300

\bibitem[{Padoan \& Nordlund(1997)}]{Padoan+Nordlund97MHD}
Padoan, P. \& Nordlund, {\AA}. 1997, astro-ph/9706176

\bibitem[{Padoan \& Nordlund(1999)}]{Padoan+Nordlund99MHD}
---. 1999, ApJ, 526, 279

\bibitem[{{Stanimirovi{\' c}} \& {Lazarian}(2001)}]{Stanimirovic+Lazarian2001}
{Stanimirovi{\' c}}, S. \& {Lazarian}, A. 2001, ApJL, 551, L53

\bibitem[{{Troland} \& {Heiles}(2001)}]{Troland+Heiles2001}
{Troland}, T.~H. \& {Heiles}, C. 2001, American Astronomical Society Meeting,
  198, 0

\end{thebibliography}









\clearpage

\clearpage

\clearpage

\end{document}